\documentclass{article}

\usepackage[final]{nips_2017}

\usepackage[utf8]{inputenc} 
\usepackage[T1]{fontenc}    
\usepackage{hyperref}       
\usepackage{url}            
\usepackage{booktabs}       
\usepackage{amsfonts}       
\usepackage{nicefrac}       
\usepackage{microtype}      
\usepackage{amsmath}
\usepackage{subcaption}
\usepackage{natbib}

\usepackage{graphicx}

\title{Predicting Future Machine Failure from Machine State Using Logistic Regression}

\author{
  Matthew Battifarano\\
  Department of Civil and Environmental Engineering\\
  Carnegie Mellon University\\
  \texttt{mbattifa@andrew.cmu.edu} \\
   \And
  David DeSmet \\
  Department of Civil and Environmental Engineering \\
  Carnegie Mellon University \\
  \texttt{ddesmet@andrew.cmu.edu} \\
  \And
  Achyuth Madabhushi\\
  Department of Civil and Environmental Engineering\\
  Carnegie Mellon University  \\
  \texttt{amadabhu@andrew.cmu.edu} \\
  \AND
  Parth Nabar \\
  Department of Civil and Environmental Engineering\\
  Carnegie Mellon University\\
  \texttt{pnabar@andrew.cmu.edu} \\
}

\begin{document}

\maketitle

\begin{abstract}
  Accurately predicting machine failures in advance can decrease maintenance cost and help allocate maintenance resources more efficiently. Logistic regression was applied to predict machine state 24 hours in the future given the current machine state.
\end{abstract}

\section{Introduction}
Understanding and predicting machine failure has become a crucial aspect of business operations. 45\% of all maintenance efforts are ineffective and 30\% of maintenance activities happen too frequently \cite{IBM-machinefailure}. A data-driven approach to predict machine failure reduces the time and cost burden associated with machine failure. Logistic Regression (LR) is a widely-used data-driven approach to binary classification. LR linearly classifies data into identifiable areas where distance from a decision boundary dictates the probability of inclusion within the class. \cite{Shalizi2012}

Prior literature has also used Support Vector Machines (SVM) to predict failures in server systems \cite{Murray2005} and hard-drives \cite{Turnbull2003}. However, SVM, unlike LR, does not provide readily interpretable outcomes that can be provided to a domain expert. 

\section{Methods}

Individual datasets were merged to construct an event stream of machine state. This event stream was used in a logistic regression model to predict the failure of the machine 24 hours in the future. The data was fit using the \texttt{LogisticRegression} model provided by the \texttt{scikit-learn} library. \cite{scikit-learn} A Jupyter notebook\cite{Kluyver2016} with the code used to generate results and figures in this report can be found on GitHub. \cite{2018HackAuton} 

\subsection{Data processing}

The data is composed of several different streams of machine events, identified by the timestamp of the observation, rounded to the hour, and an integer identifier of the machine. These datasets were merged together on observation time and machine identifier to construct an event stream of the complete machine state. Table \ref{tab:datasets} enumerates the datasets and their schemas.

\begin{table}[!ht]
\centering
\caption{Dataset schemas}
\label{tab:datasets}
\begin{tabular}{lll}
\toprule
dataset & field & description\\
\midrule
failures 	&				& event stream of component failures\\
			& machine\_id 	& unique machine identifier\\
			& datetime 		& the observation timestamp\\
            & comp\_1		& whether or not component 1 failed\\
            & comp\_2		& whether or not component 2 failed\\
            & comp\_3		& whether or not component 3 failed\\
            & comp\_4		& whether or not component 4 failed\\
\midrule
errors		&				& event stream of non-failure error codes\\
			& machine\_id 	& unique machine identifier\\
			& datetime 		& the observation timestamp\\
            & error\_1		& whether or not error 1 occurred\\
            & error\_2		& whether or not error 2 occurred\\
            & error\_3		& whether or not error 3 occurred\\
            & error\_4		& whether or not error 4 occurred\\
            & error\_5		& whether or not error 5 occurred\\
\midrule
maintenance	& 				& component maintenance records\\
			& machine\_id 	& unique machine identifier\\
			& datetime 		& the maintenance timestamp\\
            & comp\_1		& whether or not component 1 was replaced\\
            & comp\_2		& whether or not component 2 was replaced\\
            & comp\_3		& whether or not component 3 was replaced\\
            & comp\_4		& whether or not component 4 was replaced\\
            & comp\_1\_fail	& whether or not component 1 was replaced due to failure\\
            & comp\_2\_fail	& whether or not component 2 was replaced due to failure\\
            & comp\_3\_fail	& whether or not component 3 was replaced due to failure\\
            & comp\_4\_fail	& whether or not component 4 was replaced due to failure\\
\midrule
telemetry 	&				& physical measurements of the machine\\
			& machine\_id 	& unique machine identifier\\
			& datetime 		& the observation timestamp\\
            & volt			& voltage measurements\\
            & rotate		& rotation measurements\\
            & pressure		& pressure measurements\\ 
            & vibration		& vibration measurements\\
\midrule
machines	&				& description of each machine\\
			& machine\_id	& unique machine identifier\\
            & age			& age of the machine in years\\
            & model\_1		& whether or not the machine is a model 1\\
            & model\_2		& whether or not the machine is a model 2\\
            & model\_3		& whether or not the machine is a model 3\\
            & model\_4		& whether or not the machine is a model 4\\
\bottomrule
\end{tabular}
\end{table}

The event stream of the complete machine state is constructed by joining the telemetry, maintenance, and errors datasets together on machine\_id and datetime. Since telemetry contains measurements every hour for one year, event records are left joined to telemetry. The left join ensures that non-events (e.g. times at which no events occurred) are included in the event stream. The machines dataset is joined in to add age and model type to the data. Finally the day of week is extracted from the timestamp of each data point and included as a categorical feature.

Machine failure is defined as the event that at least one of its four components has failed as reported in the failures dataset. The target variable is defined as the machine failure state 24 hours after the measurement timestamp of each data point in the event steam. 

\subsection{Logistic regression}

Logistic regression is a commonly used classification method known for its easily interpretable model parameters. \cite{Agresti2003}\cite{Shalizi2012} Logistic regression models the probability of a binary target variable $Y$ given the features $X$ by mapping a linear combination of the features to $(0, 1)$ via a non-linear transformation function given by \eqref{eqn:logisitic-regression-model}.

\begin{align}
	P(Y=1 \mid X=x) = \pi(x) &= \frac{\exp(\alpha + \beta^Tx)}{1+\exp(\alpha + \beta^Tx)} \label{eqn:logisitic-regression-model}
\end{align}

In this context, each $x$ is a vector containing the complete machine state of a single machine at a single point in time and the target variable, $y$, represents the failure state of the machine 24 hours after the measurement. Explicitly, the feature vector is given by Table \ref{tab:feature-vector}. Note that machine id and timestamp are not included in the feature set, they are simply used as join criteria so that each feature contains data related to the same machine at the same time. 

\begin{table}[!ht]
\centering
\caption{The feature vector schema}
\label{tab:feature-vector}
\begin{tabular}{ll}
\toprule
feature & description\\
\midrule
error\_1		& whether or not error 1 occurred\\
error\_2		& whether or not error 2 occurred\\
error\_3		& whether or not error 3 occurred\\
error\_4		& whether or not error 4 occurred\\
error\_5		& whether or not error 5 occurred\\
comp\_1		& whether or not component 1 was replaced\\
comp\_2		& whether or not component 2 was replaced\\
comp\_3		& whether or not component 3 was replaced\\
comp\_4		& whether or not component 4 was replaced\\
comp\_1\_fail	& whether or not component 1 was replaced due to failure\\
comp\_2\_fail	& whether or not component 2 was replaced due to failure\\
comp\_3\_fail	& whether or not component 3 was replaced due to failure\\
comp\_4\_fail	& whether or not component 4 was replaced due to failure\\
volt			& voltage measurements\\
rotate		& rotation measurements\\
pressure		& pressure measurements\\ 
vibration		& vibration measurements\\
age			& age of the machine in years\\
model\_1		& whether or not the machine is a model 1\\
model\_2		& whether or not the machine is a model 2\\
model\_3		& whether or not the machine is a model 3\\
model\_4		& whether or not the machine is a model 4\\
\bottomrule
\end{tabular}
\end{table}

\subsection{Model fitting and evaluation}

Machine failure is relatively rare: only 1.7\% of samples correspond to a failure event. To compensate for the imbalance during training the model should weight the misclassification of a failure event more than misclassification of a non-failure event. To achieve this, each failure sample is given a weight of 100 and each non-failure a weight of 1. A few different values for the weight were tried, but the results appear robust to the sample weight.

To evaluate the model, 3-fold cross validation was employed. To ensure no contamination occurred between the training and testing sets, splits were constructed such that the samples in the testing set occurred chronologically after the samples in the training set and corresponded to different machines. In short, the model was evaluated on future samples from never-before-seen machines.

\section{Results}

The Logistic regression model performed well on this dataset. On held-out data the trained model was able to perfectly classify all failed events, and nearly perfectly classify all non-failure events.

\begin{figure}[!ht]
\centering
\begin{subfigure}[!t]{0.48\textwidth}
	\centering
	\includegraphics[width=2.5in]{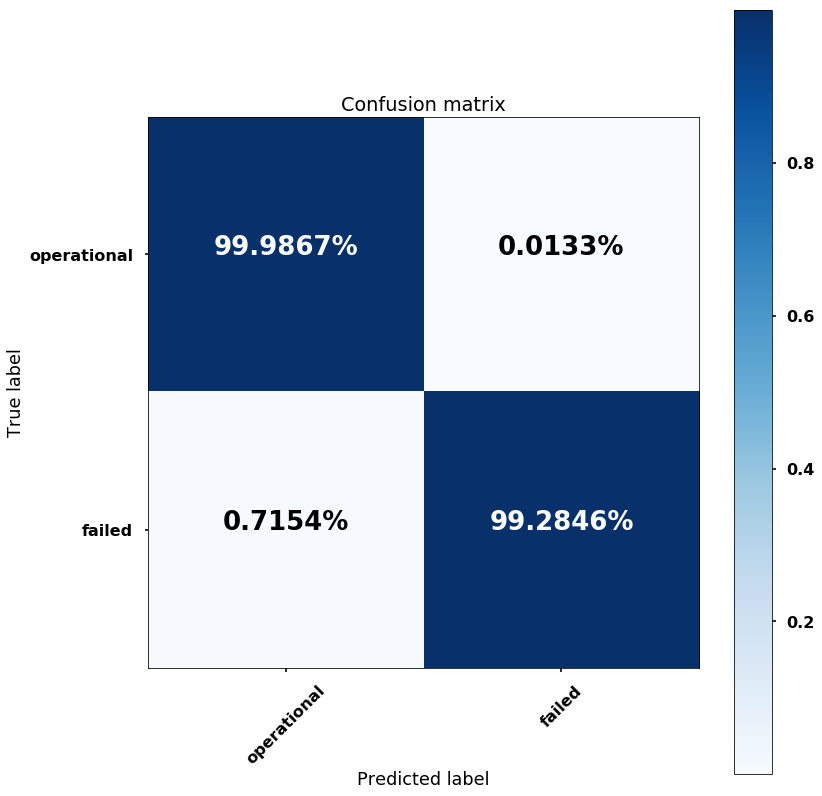}
	\caption{Normalized average confusion matrix of the validation sets of 3-fold cross validation using the full feature set.}
    \label{fig:full-features-confusion-matrix}
\end{subfigure}%
~
\begin{subfigure}[!t]{0.48\textwidth}
	\centering
	\includegraphics[width=2.5in]{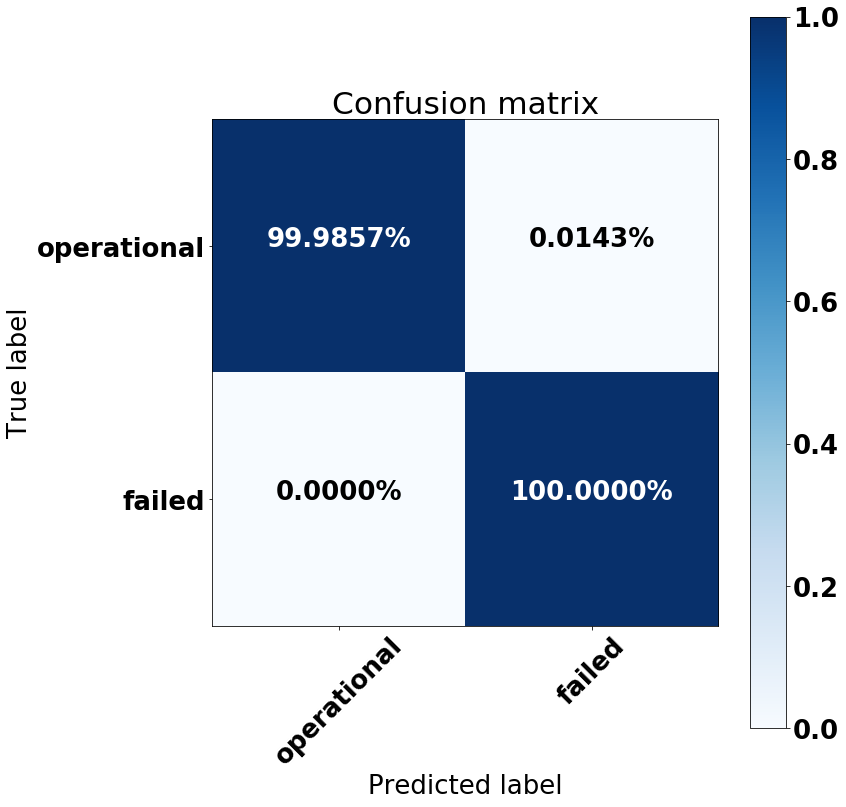}
\caption{Normalized average confusion matrix of the validation sets of 3-fold cross validation using the reduced feature set.}
\label{fig:reduced-features-confusion-matrix}
\end{subfigure}
\caption{Normalized average confusion matrices from the model trained on the full feature set (left) and the reduced feature set (right).}
\end{figure}

The initial formulation of the logistic regression was promising, however, the confusion matrix still had weight on the off-diagonal, predicting machine failures when the machine was still functional or not catching machine failures when they did occur. There was a higher percentage within the confusion matrix for false negatives\textemdash that is, when the logistic regression did not accurately predict failures when they occurred, and the team considered this quality something to be avoided.

\begin{figure}[!ht]
\centering
\includegraphics[width=5.5in]{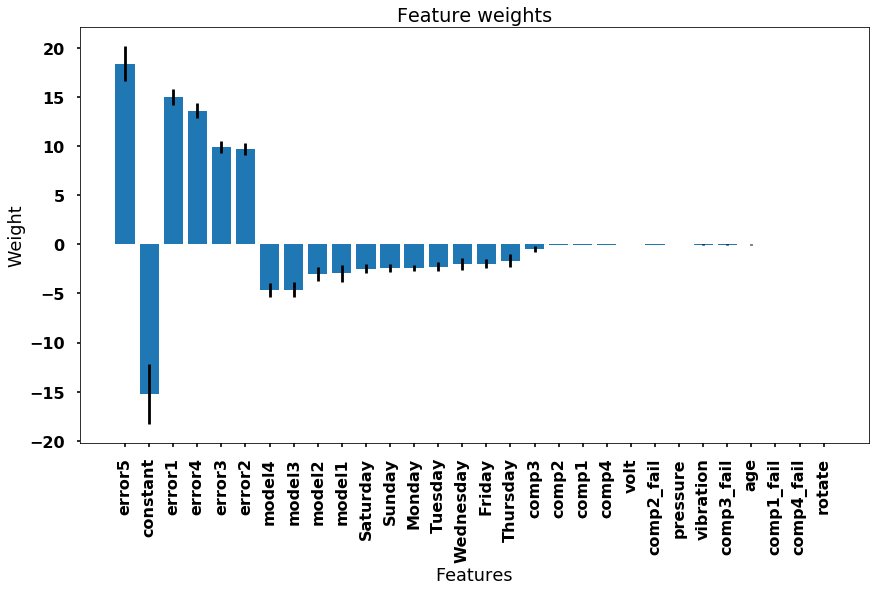}
\caption{Average values of the feature weights over 3-fold cross validation ordered by magnitude. The 'constant' feature corresponds to $\alpha$ in \eqref{eqn:logisitic-regression-model} while the rest correspond to elements of $\beta$.}
\label{fig:full-feature-weights}
\end{figure}

Additionally, an examination of the weights used in this logistic regression indicated that there was a large proportion of weights that simply were not significant to the model (see \ref{fig:full-feature-weights}). These included attribute categories like the day of the week, which component was replaced during maintenance, whether the component replace was due to failure, and the telemetry data. This indicated that these variables could be easily removed from the analysis with little consequence.

It is evident that fewer features affect machine failure than initially understood. The goal is to reduce the frequency of false-negatives by adjusting the size of the $\beta$ vector to represent fewer, but more important, features. Figure \ref{fig:reduced-features-confusion-matrix} shows the confusion matrix of the logistic regression trained on the smaller feature set. By reducing the feature set to included only the most important features, the accuracy of the logistic regression on the held-out data increased. In particular, it reduced the rate of false negatives where the machine was predicted to be operational but in fact has failed.

\section{Discussion}

While logistic regression was utilized in this analysis, there are other machine learning techniques that could be applied. For instance, Support Vector Machine classification (SVM) would have a degree of applicability. This is because SVM is easily applied to highly dimensional spaces and our dataset had 30 attributes per observation. \cite{Cortes1995}

In terms of improvement on the logistic regression, the largest potential is with false positives. That is, while demonstrating (within the test set) 100\% accuracy when it came to predicting failures, the logistic regression predicted failures in machines that were still operational. Thankfully, this is most likely where the margin of safety should be. That is, a machine failing before being caught is most likely more detrimental than a machine being inspected in a controlled environment. Nevertheless, this still is the area where our model is most in need of improvement.

One issue to explore is how sensitive logistic regression is to the time alloted before the point of failure. Currently, a 24 hour period is applied, it is interesting to attempt the logistic regression for both a shorter and longer periods. Therefore, further analysis could investigate whether or not larger periods of time would have comparable (or perhaps even improved) accuracy. Additionally, this additional analysis could incorporate a breadth of data over time.

The current logistic regression does not distinguish between components when a failure is predicted, just machine failure. Thus, further analysis might look into predicting which components will likely fail. Pinpointing component failure may further decrease maintenance costs, as the maintenance team would no longer need to have repairing capabilities for all of the components, just specific failure-prone components.

\section{Acknowledgments}
Our deepest gratitude goes to the Auton Lab at Carnegie Mellon University for providing the data set, and organizing the HackAuton. Additional thanks to the sponsors for providing the funding and datasets to make this HackAuton possible.

\bibliographystyle{plainnat}
\bibliography{sources}

\end{document}